\documentclass[twocolumn,prl,showpacs,preprintnumbers,amssymb,superscriptaddress]{revtex4}

\usepackage{epsfig}
\usepackage{dcolumn}
\usepackage{bm}

\def\preal{{\rm Re\,}}
\def\pim{{\rm Im\,}}

\def\p{\partial}
\def\a{\alpha}

\def\g{\gamma}
\def\d{\delta}
\def\beq{\begin{equation}}
\def\eeq{\end{equation}}
\def\beqa{\begin{eqnarray}}
\def\eeqa{\end{eqnarray}}
\def\Om{\Omega}

\def\oh{\frac{1}{2}}

\def\eps{\epsilon}

\def\cn{{\cal N}}
\def\cm{{\cal M}}
\def\cg{{\cal G}}
\def\G{\Gamma}

\def\raw{\rightarrow}
\def\Raw{\Rightarrow}

\def\D{\Delta}

\def\T{{\bf T}}

\begin{document}

\preprint{LMU-ASC 26/07}
\preprint{MPP-2007-49}

\title{Non-geometric fluxes as supergravity backgrounds}

\author{Fernando Marchesano}

\affiliation{ASC, Ludwig-Maximilians-Universit\"at, Theresienstra\ss e 37, 80333 M\"unchen}

\author{Waldemar Schulgin}

\affiliation{Max Planck Institut f\"ur Physik, F\"ohringer Ring 6, 80805 M\"unchen}

\begin{abstract}
We consider examples of $D=4$ string theory vacua which, although globally non-geometric, admit a local description in terms of $D=10$ supergravity backgrounds. We analyze such backgrounds and find that the supersymmetry spinors vary non-trivially along the internal manifold, reproducing the interpolating supergravity solutions found by Frey and Gra\~na. Finally, we propose a simple, local expression for non-geometric fluxes in terms of the internal spinors of the compactification.
\end{abstract}
\maketitle


Undoubtedly, fluxes have played a major role in recent attempts to connect string theory to observable physics. Why is this the case may be understood from two simple facts. First, string compactifications with background fluxes enlarge non-trivially the set of $D=4$ string vacua based on conventional Calabi-Yau compactifications,  and so the set of effective field theories that one can obtain. Second, flux compactifications are rich sources of $D=4$ effective superpotentials, allowing to address two important issues of string phenomenology, moduli stabilization and supersymmetry breaking, in a controlled manner \cite{reviews}.

A celebrated class of flux vacua is given by type IIB string theory compactified on a warped Calabi-Yau with orientifold planes \cite{drs99,gkp01}. Indeed, such constructions admit the inclusion of internal background 3-form fluxes, usually denoted by $H_3$ and $F_3$, which back-react on the Calabi-Yau geometry and produce an effective superpotential that lifts certain compactification moduli \cite{gvw99}. In many senses, one can treat the resulting $D=4$ effective theory as that obtained from the well-understood fluxless Calabi-Yau compactification, to which we add a flux-induced superpotential and a non-trivial warping. This provides a simple global picture of this class of compactifications, which even allows to treat them as an ensemble as opposed to consider each vacuum individually \cite{ddk07}. Finally, a nice feature of this set of $D=4$ vacua is that in general they can be understood in terms of $D=10$ supergravity. The geometric intuition that one obtains from this fact has triggered several phenomenologically interesting scenarios, in particular some related to early universe cosmology \cite{cosmo}.

While warped CY's have received a lot of attention, they are not the only class of flux compactifications in the literature, even if we restrict to type IIB $\cn = 1$ $D=4$ Poincar\'e invariant vacua. Another well-known family is based on compact non-K\"ahler manifolds, also containing O-planes, and dual to the solutions found in the context of heterotic strings \cite{fluxhet}. Here the set of fluxes not only consists of field strengths like $F_3$, but also of torsion factors that deviate the metric from the Calabi-Yau one, and so they are usually referred to as geometric fluxes \cite{glmw02,kstt02}.

While geometric fluxes are less intuitive than the usual ones, an even more exotic kind of fluxes are those named non-geometric fluxes, which arise in the context of non-geometric compactifications \cite{dh02hmw02fww04,Hull04,stw05,dh05,acfi06}. Such compactifications are possible because, as string theory has a larger duality group than plain supergravity, it can be compactified  in spaces where (classical) gravity would not even make sense. Perhaps the best understood class of non-geometric vacua are those based on T-folds \cite{Hull04}, where locally there is a standard spacetime description of the compactification but, as we relate two overlapping patches, we must use  transition functions that mix the metric with the antisymmetric B-field and interchange KK with winding modes, just as T-dualities do.

Clearly, non-geometric compactifications are a novel and exciting area where to develop string phenomenology, as they are genuine string compactifications that cannot be described as $D=10$ supergravity plus localized sources. Moreover, as non-geometric fluxes are combined with the previous fluxes, the effective superpotential that one obtains generalizes that of a Calabi-Yau with fluxes \cite{stw05}, in such a way that compactifications with all moduli stabilized are easily obtained \cite{allmoduli}.

On the other hand, our current understanding of non-geometric compactifications is still too weak in order to perform a detailed analysis of their phenomenological possibilities. Part of the problem is that, unlike for other fluxes, there is no simple, intuitive definition of what non-geometric fluxes mean. In practice, they appear as the structure constants of $D=4$ effective gauged supergravities, or as internal  monodromies between the metric and B-field \cite{kstt02,stw05,dh05}. In fact, very few examples of non-geometric compactifications have been built beyond Scherk-Schwarz compactifications and close relatives and, because of the issues above, it is hard to figure out how the space of non-geometric vacua looks like.

The purpose of this note is to shed further light into our understanding of non-geometric fluxes by describing them in terms of supergravity backgrounds. More precisely, we focus on some simple examples of T-folds, where a local $D=10$ description is possible, and study in detail the supergravity solutions associated to them. This will not only provide us with a simple definition of non-geometric flux, but also unveil many local properties of this kind of backgrounds. Notice that the latter point is essential in order to build realistic scenarios from non-geometric compactifications. Indeed, most of the cosmological models constructed from warped CY's with fluxes are based on D-brane inflation, where only a local description of the supergravity background (like, e.g., the KS throat \cite{ks00}) is needed. Hence, based on the results below, similar scenarios could presumably be constructed for non-geometric vacua.

In order to perform our analysis, let us first construct a non-geometric vacuum. As pointed out in \cite{kstt02} this can be done by performing appropriate T-dualities on simple geometric flux compactifications. In particular, let us consider type IIB string theory compactified on the warped toroidal background \footnote{We are using the convention $4\pi^2\a' = 1$.}
\beqa
\label{fullmetric1}
ds^2 & = & Z^{-1/2} ds^2_{M_4} +  Z^{1/2} ds^2_{{\T^6}}\\
\label{metric1}
ds^2_{\T^6} & = & \sum_{i=1}^3 ds^2_{(\T^2)_i},  \, ds^2_{(\T^2)_i}\, =\,  R_i^2 |dx^i + \tau^i\, dx^{i+3}|^2 \\
\label{H31}
H_3 & = & N\, (dx^1 \wedge dx^5 - dx^4 \wedge dx^2) \wedge dx^6 \\
\label{F31}
F_3 & = & N\, (dx^4 \wedge dx^2 - dx^1 \wedge dx^5) \wedge dx^3 \\
\label{F51}
\tilde{F}_5 & = & (1 + *_{10})\, d{\rm vol}_{M_4} \wedge dh \\
\label{dilaton1}
\tau & = & C_0 + ie^{-\phi_0}\, =\, const. 
\eeqa
which is closely related to the vacua considered in \cite{kstt02}, and a simple example of a warped Calabi-Yau with $H_3$ and $F_3$ fluxes of \cite{gkp01}. Here $ds^2_{M_4}$ stands for the $D=4$ Minkowski metric, $\tau$ for the type IIB axio-dilaton, the warp factor $Z$ only depends on the internal coordinates $x^i \sim x^i  +1$, $N$ is an even integer \cite{fp02}, and we must impose $dh = e^{-\phi_0} dZ^{-1}$ \cite{gkp01}. In order to find a solution to the Bianchi identity of $\tilde{F}_5$ and Einstein's equations we need to add negative tension objects to the compactification \cite{gkp01}, which in our case will be 64 O3-planes expanding $M_4$. Finally, by adding $32 -2N^2$ D3-branes and no further localized objects all consistency conditions will be satisfied.

The presence of the background fluxes $H_3$ and $F_3$ generates a superpotential \cite{gvw99}
\beq
W\, =\, \int_{\T^6} (F_3 - \tau H_3) \wedge \Om_3 \, = \,  N\, (\tau^1 - \tau^2) (\tau^3 - \tau)
\label{super}
\eeq
and so we also need to impose $\tau^1 = \tau^2$ and $\tau^3 = \tau$. This will guarantee that the complexified 3-form flux $G_3 = F_3 -\tau H_3$ is an imaginary-self-dual (ISD), primitive (2,1)-form, and so we will have a supersymmetric vacuum \cite{gp00}. In fact, due to the simplicity of our example, the present compactification yields a $D=4$ $\cn = 2$ effective theory. The supersymmetry generators are of the form
\beqa
\label{spinorL}
\eps_L & = & \frac{1}{\sqrt{2}} \left(u \otimes \chi_L + u^* \otimes \chi^*_L \right) \\
\label{spinorR}
\eps_R & = & \frac{1}{\sqrt{2}} \left(u \otimes \chi_R + u^* \otimes \chi^*_R \right)
\eeqa
where $u$ stands for the external and $\chi$ for the internal spacetime spinor, the latter chosen of negative chirality. While the above ansatz is valid for general supersymmetric type IIB compactifications, here the presence of the flux $G_3$ and the O3-planes imposes the following relation
\beq
\eps_L\,  =\,  \G_{O3}\, \eps_R \, = \, -i \G^{(6)}\, \eps_R
\label{pO3}
\eeq
between left ($\eps_L$) and right ($\eps_R$) spinors. Here $\G^{(6)}$ stands for the internal chirality operator, and so we have that
\beq
\chi_L = i \chi_R \quad \raw \quad \eps = \eps_L + i \eps_R = \sqrt{2}\, u \otimes \chi_L
\label{typeB}
\eeq
which is the usual B-type ansatz for warped Calabi-Yau compactifications with fluxes \cite{bb96}. In the present background, the internal spinors are of the form $\chi_{R,i} = Z^{-1/8} \eta_i$, where  
\beq
\eta_1\, =\, \oh (---),\quad \eta_2 \, =\, \oh (++-)
\label{spinors1}
\eeq
in the usual spinor notation.

As pointed out in \cite{kstt02}, by performing two consecutive T-dualities along two directions of a $H_3$ component one obtains a non-geometric compactification. In particular, following the notation of \cite{stw05}, after two T-dualities along $\{x^5, x^1\}$ the flux $H_3 = dB$ partially becomes a non-geometric Q-flux via
\beq
H_{x^1x^5x^6} = N \quad \stackrel{T_{x^5x^1}}{\longrightarrow} \quad Q_{x^6}^{x^5x^1} = N
\label{Qflux}
\eeq
whereas the other component of $H_3$ remains untouched.

In order to obtain the new background one can apply Buscher's T-duality rules \cite{Buscher87}, but first one needs have the directions $\{x^5, x^1\}$ as isometries of the background. This is achieved by choosing an appropriate gauge for the B-field, like $B = N x^6 (dx^1 \wedge dx^5 - dx^4 \wedge dx^2)$, and by smoothing out the localized sources (i.e., the D3-branes and O3-planes) along $\{x^5, x^1\}$. This means in particular that the warp factor $Z$ will be given by a function $\D \equiv \D(x^2,x^3,x^4,x^6)$, which by Einstein's equations must satisfy
\beq
- e^{-\phi_0} \nabla^2_{\T^4} \D\, =\,  \frac{2N^2}{{\rm Vol}_{\T^4}} + \sum_{i} q_i \d_{\T^4}(\vec{x}_{i}) 
\label{warpeq}
\eeq
where $\T^4$ stands for the four-torus expanded by $\{x^2,x^4,x^3,x^6\}$ and the delta functions represent the sources localized in the same $\T^4$, with a charge $q = 1$ in the case of a D3-brane and $q = -2$ for a (partially delocalized) O3-plane. Notice that on $\T^4$ the O3-planes are at the fixed locations $x^i \in \mathbb{Z}/2$ ($i = 2, 3, 4, 6$), and in the following we will focus our attention to the neighborhood containing the one at the origin. Finally, in order to simplify the discussion below, we will choose the compactification moduli to be $R_1 = R_2 = 1$ and $\tau^1 = \tau^2 = i$, although more general choices consistent with the supersymmetry condition $\tau^1 = \tau^2$ may also be chosen. 

Taken into account these prescriptions and directly applying Buscher's rules along $\{x^5, x^1\}$ one again obtains type IIB string theory, but this time on the background
\beqa
\label{fullmetric2}
ds^2 & = & \D^{-1/2} ds^2_{M_4} +  \D^{1/2} ds^2_{{\cm_6}}\\
\label{metric2}
ds^2_{\cm_6} & = & (dx^2)^2 + (dx^4)^2 +  dz^3 d\bar{z}^3 \\
\nonumber
& & +\, e^{2(\phi - \phi_0)} \left((dx^1)^2 + (dx^5)^2\right)  \\
\label{B22}
B & = & N x^6\, (dx^2 \wedge dx^4 + e^{2(\phi -\phi_0)} dx^5 \wedge dx^1) \\
\label{F12}
F_1 & = & N\,\preal dz^3\\
\label{F32}
\tilde{F}_3 & = & F_1 \wedge B + e^{2\phi-\phi_0} *_{\cm_6} d e^{-2\phi} \wedge dx^5 \wedge dx^1  \\
\label{F52}
\tilde{F}_5 & = & (1 + *_{10})\, d{\rm vol}_{M_4} \wedge d (N x^6/\D) e^{-\phi_0}\\
\label{dilaton2}
e^{\phi - \phi_0} & = & \left(\D + (Nx^6)^2\right)^{-1/2}
\eeqa
where for simplicity we have defined $dz^3 = dx^3 + \tau dx^6$, with $\tau$ given by (\ref{dilaton1}). When compactifying this theory, one should perform the identifications $x^3 \sim x^3 + 1$ and $x^6 \sim x^6 +1$. However, for reasons that will become clear below, we will initially not impose such identifications. On the other hand, we will still impose $x^i \sim x^i + 1$ for $i = 1,2,4,5$. The background fluxes $\tilde{F}_3 = F_3 - C_0 H_3$, $F_1 = dC_0$ stand for the generalized field strength of type IIB supergravity. Notice that neither the axion $C_0$ nor the dilaton $e^{-\phi}$ are constant in this new background. Nevertheless, as usual one can still define an average string coupling by setting $g_s = e^{\phi_0}$, with $e^{\phi_0}$ given by (\ref{dilaton1}), that can be taken to arbitrary small values.

Just as the above background has been T-dualized we must also transform the localized sources, namely the orientifold planes and the open string sector of the theory. By the usual T-duality rules we would now expect to have O5-planes and D5-branes wrapped on the two-torus expanded by $\{x^5, x^1\}$. This intuition is confirmed by looking at the Bianchi identities of the new background. While $dF_1 = dH_3 = 0$ are clearly satisfied, for $\tilde{F}_3$ we have that \footnote{Following the Hodge star conventions in \cite{gkp01}, for a $p$-form $\a$ on a manifold $\cm$ we have that $\a \wedge *_{\cm} \a = - \a\cdot\a\, d{\rm vol}_\cm$.}
\beqa
\nonumber
d\tilde{F}_3 - H_3 \wedge F_1  & = & g_s^{-1} \nabla^2_{\T^4} e^{-2 (\phi -\phi_0)}\, d{\rm vol}_{\T_4} \\
\label{BIF32}
& = & - \sum_{i} q_i \d_{\T^4}(x_{i})\, d{\rm vol}_{\T_4}
\eeqa
where we have used (\ref{F32}) and (\ref{warpeq}), and we are defining $d{\rm vol}_{\T^4} = R_3^2 \pim \tau\, dx^2 \wedge dx^4 \wedge dx^3 \wedge dx^6$. Hence we have a set of localized sources to which the RR field strength $F_3$ couples. As we expected, these are D5-branes and O5-planes, expanding $M_4 \times \T^2_{(x^5, x^1)}$ and being pointlike in $\{x^2, x^4, x^3, x^6\}$. In particular, there will be an O5-plane at the origin $x^2 = x^3 = x^4 = x^6 = 0$ and, if we treat such coordinates as non-compact, one at each location given by $x^i \in \mathbb{Z}/2$ ($i = 2, 3, 4, 6$), as we would also expect from T-duality. The remaining Bianchi identity can be easily computed if we realize that the internal component of $\tilde{F}_5$ satisfies
\beq
\tilde{F}_5 - B \wedge \tilde{F}_3 + \oh B^2 \wedge F_1 = -  e^{-2(\phi-\phi_0)} N *_{\cm_6} \pim d z^3
\label{relF52}
\eeq
and so the r.h.s. of (\ref{relF52}) is closed. This implies that
\beq
d\tilde{F}_5 - H_3 \wedge \tilde{F}_3  =  B \wedge \left( d\tilde{F}_3 -H_3 \wedge F_1 \right)
\label{BIF52}
\eeq
and so, by (\ref{BIF32}), we again have sum of delta functions. The presence of the $B$-field is easy to understand from the fact that the l.h.s. of (\ref{BIF52}) measures the D3-brane charge of localized sources, and this same charge is induced on D5-branes in the presence of a non-trivial B-field \cite{Douglas95}. Hence the r.h.s. of (\ref{BIF52}) takes into account that we are in the presence of D5-branes magnetized by a non-trivial B-field. What is perhaps more surprising is the fact that the O5-planes which are not located at $x^6 = 0$ also seem to be `magnetized'. In that case we should be dealing with exotic orientifold planes, analogous to the ones analyzed in \cite{adm02}. Finally, in the same way that we have solved the Bianchi identities, one can check that the equations of motion for the metric, axio-dilaton and background fluxes are satisfied. As a result, even if one is skeptical about applying Buscher's on the initial type IIB background, it is clear that eqs.(\ref{fullmetric2})-(\ref{dilaton2}) describe a genuine type IIB supergravity vacuum.

Notice that in this new background the warp factor $Z'$ and the 4-form potential $C_4' = h' d{\rm vol}_{M_4}$ are given by
\beq
Z'\, =\, \D,\quad h'\, =\, Nx^6 e^{-\phi_0} \D^{-1}
\label{newforms}
\eeq
and so the usual relation $d(h - e^{-\phi}Z^{-1})=0$ of type IIB compactifications with ISD $G_3$ fluxes, D3-branes and O3-planes is no longer satisfied. Such relation is only satisfied in the limit $x^6 \raw \infty$, where the $B$-field is so strong and the two-torus $\{x^5, x^1\}$ so small that a D5-brane wrapped on it looks like a D3-brane. On the other hand, on the limit $x^6 \raw - \infty$ we recover the relation $d(h + e^{-\phi}Z^{-1})=0$, satisfied by type IIB backgrounds with IASD $G_3$ fluxes, anti-D3-branes and anti-O3-planes. Finally, at $x^6 = 0$ we have that $h=0$ and $Z = e^{-2(\phi- \phi_0)}$, which is typical of type IIB flux compactifications containing O5-planes and geometric fluxes.

It thus seems that, as we move along $x^6$, our type IIB background interpolates between different classes of type IIB compactifications. We can parameterize such interpolation by defining an angle $\a$ as
\beq
{\rm sin}\, \a \, =\, \D^{1/2} e^{\phi -\phi_0},\quad {\rm cos}\, \a \, =\, Nx^6 e^{\phi -\phi_0}
\label{defangle}
\eeq
and analyzing how does the background depend on it. A non-trivial dependence will appear on the forms $J$ and $\Om$ describing the metric, which can be taken to be
\beqa
\nonumber
\frac{J}{\D^{1/2}} & = & e^{2(\phi-\phi_0)} dx^5\wedge dx^1 + dx^2 \wedge dx^4  +  i\frac{R_3^2}{2} dz^3 \wedge d\bar{z}^3\\
\nonumber
\frac{\Om}{\D^{3/4}} & = & i e^{\phi-\phi_0} R_3\, (dx^1 +  idx^5) \wedge (dx^4 + i dx^2) \wedge dz^3
\eeqa
which naturally define a complex and a symplectic structure. In terms of these definitions, one can check that the ISD, (2,1)-form in this background is
\beq
\nonumber
\cg_3  =  \tilde{F}_3 - i e^{-\phi}\, \preal \left(e^{i\a} \left[H_3 - i \D^{1/2} d(\D^{-1/2} J) \right] \right) 
\label{inter3formflux}
\eeq
which, at specific values of $\a$ takes a more familiar form
\beqa
\label{limitinf2}
\a \raw 0 & \Raw & \cg_3 =  \tilde{F}_3 - i e^{-\phi} H_3  = G_3\\
\label{limit02}
\a = \pi/2 & \Raw & \cg_3 = \tilde{F}_3 - i e^{-2\phi} d(e^{\phi} J) \\
\label{limit-inf2}
\a \raw \pi & \Raw & \cg_3 =  \tilde{F}_3 + i e^{-\phi} H_3 \, =\, \bar{G}_3
\eeqa
Finally, a D5-brane wrapped on $\{x^5, x^1\}$ satisfies the conditions
\beq
\iota_X \Om|_{D5}  = \pim \left(e^{-i\a} (B+iJ)\right)|_{D5}  =  0
\label{susyD5}
\eeq
which are the usual BPS conditions for a D5-brane in a Calabi-Yau compactification \cite{mmms99}, notice however that in the Calabi-Yau case $\a$ is a constant parameter, while here it varies along the compactification manifold.

But the easiest way to see that this supergravity background interpolates between different classes of standard type IIB compactifications is by analyzing the supersymmetry spinors. These spinors can also be obtained by T-duality, by applying the rules of \cite{Hassan99}. One then obtains that the new spinors no longer satisfy the relation (\ref{typeB}) but instead 
\beq
\eps_L'  = \left({\rm cos}\, \a \, \G_{O3} + {\rm sin}\, \a\, \G_{O5} \right)\, \eps_R'
\label{typeBC}
\eeq
where $\G_{O5}$ is the chirality operator on the coordinates $\{x^2, x^4, x^3, x^6\}$ transverse to $M_4 \times \T^2_{(x^5, x^1)}$, as well as the orientifold projection that an O5-plane wrapping $M_4 \times \{x^5, x^1\}$ would implement in standard compactifications.

Now, since our initial background contained two independent spinors so will the new one. In order to see how the $D=10$ spinors (\ref{spinorL}) and (\ref{spinorR}) now look like, let us consider a particular linear combination of them. Namely, we choose
\beq
\label{killingnew}
\chi_{R,\pm}\,  =\, \D^{-1/8}  \eta_\pm, \quad  \eta_\pm\, =\,  \frac{1}{\sqrt{2}} (\eta_1 \pm \eta_2) \\
\eeq
then, taking into account that
\beqa
\label{actionO3}
\G_{O3}\, \eta_\pm\, =\, i \eta_\pm & & \G_{O3}\, \eta_\pm^*\, =\, - i \eta_\pm^* \\
\label{actionO5}
\G_{O5}\, \eta_\pm\, =\, \pm \eta_\pm & & \G_{O5}\, \eta_\pm^*\, =\, \pm \eta_\pm^*
\eeqa
we obtain the following $D=10$ spinors
\beqa
\label{spinorRnew}
\eps_{R,\pm} & = & \frac{1}{\sqrt{2}} \left(u \otimes \chi_{R,\pm}  + u^* \otimes \chi^*_{R,\pm} \right) \\
\label{spinorLnew}
\eps_{L,\pm} & = & \frac{1}{\sqrt{2}} \left( i\, e^{\pm i\a} u \otimes \chi_{R,\pm} -i\, e^{\mp i\a}  u^* \otimes \chi^*_{R,\pm} \right)
\eeqa
precisely matching the spinor ansatz considered in \cite{fg03}, where type IIB supergravity solutions interpolating between standard compactification ans\"atze were shown to be possible.

In particular, the interpolating solutions analyzed in \cite{fg03} connected backgrounds where D3-branes are BPS objects to those where D5-branes are BPS. We see that, if we extend our supergravity solution along $x^6 \in (-\infty,\infty)$ the same situation will happen, with the difference that now anti-D3-branes will also be included. So, just as expected, our supergravity background is such that for $x^6 \raw -\infty$ anti-D3-branes are BPS, for $x^6 = 0$ D5-branes are BPS and for $x^6 \raw \infty$ D3-branes are BPS. At intermediate points of $x^6$ the supersymmetry preserved by the background is that of a magnetized D5-brane.

This description fits exactly with the D5-brane BPS conditions found in (\ref{susyD5}), which can be derived as follows. First, one follows the computations of \cite{fg03} in order to see that (\ref{spinorLnew}) and (\ref{spinorRnew}) are indeed supersymmetry generators of the new background. Second, one chooses one of the two spinors, say $\eta_+$, and constructs the spinor bilinears
\beq
J_{mn}\, = \, - i \eta_+^\dag {\G}_{mn} \eta_+,\quad \Om_{mnp}\, = \, - \eta_+^T {\G}_{mnp} \eta_+
\label{bilinears}
\eeq
familiar from $SU(3)$-structure compactifications \cite{reviews}, and obtains the same forms $J$ and $\Om$ as above. Finally, by repeating the $\kappa$-symmetry computations of \cite{mmms99} one arrives at (\ref{susyD5}). Had we chosen the internal spinor $\eta_-$, we would have obtained a different two-form $J$, but nevertheless the same final condition for a D5-brane wrapped on $\{x^5, x^1\}$. Alternatively, the same conclusion can be reached by using the results in \cite{ms05}.

To sum up, we have analyzed the type IIB supergravity backgrounds associated to a simple set of non-geometric backgrounds, and found that they correspond to solutions that interpolate between standard type IIB compactifications. This interpolation is particularly manifest in the internal part of the space-time supersymmetry generators, which fit into the ansatz used in \cite{fg03}. Although our background is particularly simple and yields $D=4$ extended supersymmetry, we do not expect this to be the general case. In fact, by analyzing other simple non-geometric backgrounds we find that the ansatz (\ref{spinorRnew}) and (\ref{spinorLnew}) is not always realized, but rather more general interpolating ans\"atze like that used in \cite{Dall'agata04} and generalizations. What does remain valid is the relation (\ref{typeBC}) between left and right-handed spinors, in the sense that we have a rotation between two usual  orientifold projections $\G_{Op}$ and $\G_{Oq}$ with a rotation parameter $\a$ varying along the internal manifold.
This should not only be true for toroidal-like non-geometric vacua, but also for more general ones that can be obtained from fiberwise T-duality on a geometric flux background. It would be interesting to verify this by generalizing the computations performed in \cite{fmt03}. Remarkably, the rotation ansatz (\ref{typeBC}) is compatible with the presence of orientifold planes, as our example explicitly shows. Indeed, by construction we know that there is an O5-plane located at $x^6 = 0$. At this point $\a = \pi/2$ and so (\ref{typeBC}) reduces to the usual O5-plane projection (in flat space), but notice that this is no longer true as soon as $x^6 \neq 0$.

When describing this background as a non-geometric compactification, one considers $\{ x^2, x^4, x^3, x^6\}$ to expand a four-torus $\T^4$, over which the $\T^2$ expanded by $\{x^5, x^1\}$ is fibered. As we perform the identification $x^6 \sim x^6 + 1$, $\T^2_{(x^5, x^1)}$ will suffer a non-geometric monodromy (namely a T-duality transformation) that signals the presence of the non-geometric Q-flux (\ref{Qflux}). Notice that this is a global definition, and that there is no obvious local definition of a Q-flux, like the one for a 3-form flux $H_3$ via $H_3 = dB$. However, our background example above suggests how such definition could be. Consider the quantity
\beq
q_{a}^{b c}\, = \, \p_{a} \left( { \chi_R^\dagger\, \g^{bc}\, \G_{Op}\, \chi_L \over e^{\phi -\phi_0}\, \chi_L^\dagger \chi_L}\right)
\label{localdef}
\eeq
defined in a local, geometric neighborhood containing an $Op$-plane, where $\g^{bc}$ is the antisymmetrized product of two $\g$-matrices. The usual Q-flux is then obtained by integrating (\ref{localdef}) over a one-cycle $\g_a$:
\beq
Q_{a}^{bc}\, =\, \int_{\g_a}\, q_a^{bc}
\label{globaldef}
\eeq
(see \cite{defs} for alternative, possibly related definitions). It is easy to see that, by plugging (\ref{defangle}) and (\ref{typeBC}) in the above expressions and setting $p = 5$ one recovers the Q-flux component (\ref{Qflux}) present in our background.

It would be interesting to compute (\ref{localdef}) and (\ref{globaldef}) for more involved non-geometric compactifications, as well as to generalize the notion of Q-charge, along the lines of \cite{gh05,glw06,dft07}. In any case, it should be clear what the intuition behind the above definitions is. The local quantity (\ref{localdef}) measures how the relative rotation generated by $\g^{bc}$ between left and right spinors varies along the internal manifold. Hence it will always vanish for the standard, geometric ans\"atze \cite{fluxhet,bb96}. On the other hand, if (\ref{globaldef}) does not vanish for some closed path $\g_a$, then there is a spin monodromy that acts differently for the internal spinors $\chi_L$ and $\chi_R$. This could not possibly be for a geometric compactification, since in that case both $\chi_L$ and $\chi_R$ are the same kind of objects, i.e., sections of the same spin bundle, and should transform in the same way when going around $\g_a$. Thus, (\ref{localdef}) and (\ref{globaldef}) provide a suitable way to measure global non-geometrical aspects of string compactifications. The hope is that, with these definitions at hand, one can better understand what the space of non-geometric vacua is.

\vspace*{.75cm}

\centerline{\bf Acknowledgements}

\vspace*{.25cm}

We wish to thank A.~Font, M.~Gra\~na, L.~E.~Ib\'a\~nez and P.~Koerber for useful discussions.
The work of F.M. is supported by the European Network ``Constituents, Fundamental Forces and Symmetries of the Universe", under the contract MRTN-CT-2004-005104. W.S. would like to thank the University of Munich for hospitality.

\end{document}